\documentclass[
aps,%
12pt,%
final,%
notitlepage,%
oneside,%
onecolumn,%
nobibnotes,%
nofootinbib,% 
superscriptaddress,%
noshowpacs,%
centertags]%
{revtex4}
\usepackage{amsmath, amsthm, amssymb}
\usepackage[all]{xy}
\usepackage{multirow}

\newcommand{\Dl}{\Delta}
\newcommand{\g}{\gamma} %gamma

\newcommand{\la}{\lambda} %lambda

\newcommand{\sg}{\sigma} %sigma

\newcommand{\Ccal}{\mathcal{C}}

\newcommand{\Ncal}{\mathcal{N}}

\newcommand{\Pcal}{\mathcal{P}}

\newcommand{\Sbb}{\mathbb{S}}
\newcommand{\Zbb}{\mathbb{Z}}

\newcommand{\lp}{\left(}
\newcommand{\rp}{\right)}
\newcommand{\lc}{\left\{}
\newcommand{\rc}{\right\}}

\newcommand{\de}{{\rm d}}

\newcommand{\w}{\!\wedge\!}

\newcommand{\RP}{\mathbb{RP}^{2}} %RP^2

\newcommand{\tx}{\text}
\newcommand{\qch}{\text{\boldmath $q$}}

\begin{document}

\selectlanguage{english}

\begin{flushright} % Paper number
{\small OU-HET 876} \\[-.35em]
{\small RIKEN-STAMP-20}
\end{flushright}

\title{Single-flavor Abelian mirror symmetry on $\RP \times \Sbb^{1}$} % Title

% Author 1
\author{\firstname{Hironori}~\surname{Mori}}
\email{hiromori@het.phys.sci.osaka-u.ac.jp}
\affiliation{Department of Physics, Graduate School of Science, Osaka University}
%\address{Toyonaka, Osaka 560-0043, Japan}
\thanks{The work of H.M. was supported in part by the JSPS Research Fellowship for Young Scientists}

% Author 2
\author{\firstname{Takeshi}~\surname{Morita}}
\email{t-morita@cr.math.aci.osaka-u.ac.jp}
\affiliation{Graduate School of Information Science and Technology, Osaka University}
%\address{Toyonaka, Osaka 560-0043, Japan}
%\thanks{The work of T.M. was supported in part by the JSPS Research Fellowship for Young Scientists}

% Author 3
\author{\firstname{Akinori}~\surname{Tanaka}}
\email{akinori.tanaka@riken.jp}
\affiliation{iTHES Research Group, RIKEN}
%\address{Wako, Saitama 351-0198, Japan}
\thanks{This research was supported by the RIKEN iTHES Project.}

\begin{abstract} % Abstract
The supercoonformal index on $\RP \times \Sbb^{1}$ can be derived exactly by the localization technique and applied  to the direct proof of Abelian mirror symmetry. We find two sets of parity conditions compatible with the unorientable property of $\RP$ and then rigorously show two kinds of Abelian mirror symmetry via the index on $\RP \times \Sbb^{1}$.
%The supercoonformal index defined as the path integral over $\RP \times \Sbb^{1}$ can be derived exactly by using the localization technique and applied  to directly proving Abelian mirror symmetry on this unorientable manifold. We classify parity conditions which should be imposed on the fields to be consistent with the unorientable property of $\RP$. It is turned out that there exist two allowed conditions for a vector multiplet which we name $\Pcal$- and $\Ccal\Pcal$-type, and accordingly, we show two kinds of mirror symmetry in terms of the index on $\RP \times \Sbb^{1}$.
\end{abstract}

\maketitle

%%%%%%%%%%%%%%%%%% section 1 %%%%%%%%%%%%%%%%%%%%%%%%%%%%%%%%%
\section{Introduction}
An interesting observation in three-dimensional supersymmetric field theories is Abelian mirror symmetry \cite{Intriligator:1996ex, Aharony:1997bx, Kapustin:1999ha}, whose simplest version holds between the supersymmetric QED (SQED) and the so-called XYZ model. This duality could be proven in terms of exact physical quantities \cite{Hama:2010av, Imamura:2011su, Krattenthaler:2011da, Kapustin:2011jm} computed by the localization \cite{Pestun:2007rz}.
%This duality on compact curved manifolds could be exactly proven \cite{Hama:2010av, Imamura:2011su, Krattenthaler:2011da, Kapustin:2011jm} in terms of the partition functions and the superconformal indices (SCIs) computed by the recent development of the localization \cite{Pestun:2007rz}.
In \cite{Tanaka:2014oda, Tanaka:2015pwa}, we derive the superconformal index (SCI) on an \textit{unorientable} manifold $\RP \times \Sbb^{1}$, which encodes the parity conditions imposed on the fields to be consistent with the unorientable structure. We name these $\Pcal$-type and $\Ccal\Pcal$-type.
%Those names are originated from the fact that they accompany the corresponding parity action for a matter multiplet with some charge $\gch$ under the gauge symmetry.
As a result, we can realise two new types of single-flavor Abelian mirror symmetry on $\RP \times \Sbb^{1}$ as mathematically quite non-trivial identities of the SCIs. %depending on if the dynamical gauge fields in the SQED is set to be $\Pcal$- or $\Ccal\Pcal$-type one.

%%%%%%%%%%%%%%%%%% section 2 %%%%%%%%%%%%%%%%%%%%%%%%%%%%%%%%%
\section{$\Ncal = 2$ quantum field theories on $\RP \times \Sbb^1$}

%%%%%%%%%%%%%%%%%% section 2.1 %%%%%%%%%%%%%%%%%%%%%%%%%%%%%%%%
\subsection{Parity conditions}
The unorientable manifold $\RP$ is constructed by the antipodal identification acting on a two-sphere $\Sbb^2$ by $( \vartheta, \varphi, y ) \sim ( \pi - \vartheta, \pi + \varphi, y )$, where $\vartheta \in [ 0, \pi ]$, $\varphi \in [ 0, 2 \pi ]$ on $\Sbb^2$, and $y \in [ 0, 2 \pi ]$ along $\Sbb^1$ 3rd direction. In 3d $\Ncal = 2$ theories, the parity conditions under such identification have to be set on a vector multiplet $V = ( A_{\mu}, \sg, \la, \overline{\la}, D )$ and a matter multiplet $\Phi = ( \phi, \psi, F )$ to satisfy some physical requirements. It is turned out that the conditions for $V$ are classified as $\Pcal$-type and $\Ccal\Pcal$-type. The latter means that its action accompanies charge conjugation $\Ccal$ with parity $\Pcal$, and this consequence leads to the phenomenon that two matters $( \phi_{1}, \phi_{2} )$ with the opposite gauge charges $( + \qch, - \qch )$, respectively, are merged into a doublet field $\Phi_{\text{d}}$, which means $\Phi_{\text{d}}$ should be treated as a degree of freedom on $\Sbb^2$. We summarise them in \eqref{p} and \eqref{cp} (the arrow indicates the action of the antipodal identification).
\begin{align}
% P-type
	&
	\begin{tabular}{|c|l|c|} \hline \label{p}
	& \multicolumn{1}{c|}{Vector multiplet $V^{( \Pcal )}$} & Matter multiplet $\Phi_{\text{s}}$ \\ \hline \hline %1
	\multirow{4}{*}{ $\Pcal$-type } & $A_{\vartheta} \to - A_{\vartheta}$, $A_{\varphi, y} \to + A_{\varphi, y}$, & \multirow{3}{*}{$\begin{aligned} \phi &\to \phi, \\ \psi &\to - i \g_{1} \psi, \\ F &\to F, \end{aligned}$} \\ %2
	& $\sg \to - \sg$, & \\ %3
	& $\la \to + i \g_{1} \la$, $\overline{\la} \to - i \g_{1} \overline{\la}$, & \\ %4
	& $D \to + D$. & with a gauge charge $\qch$. \\ \hline %5
	\end{tabular} \\[.75em]
% CP-type
	&
	\begin{tabular}{|c|l|l|} \hline \label{cp}
	& \multicolumn{1}{c|}{Vector multiplet $V^{( \Ccal\Pcal )}$} & \multicolumn{1}{c|}{Matter multiplet $\Phi_{\text{d}}$} \\ \hline  \hline %1
	\multirow{4}{*}{$\Ccal\Pcal$-type} & $A_{\vartheta} \to + A_{\vartheta}$, $A_{\varphi, y} \to - A_{\varphi, y}$, & \multirow{2}{*}{${\scriptstyle \begin{pmatrix} \phi_{1} \\ \phi_{2} \end{pmatrix} } \to {\scriptstyle \begin{pmatrix} \phi_{2} \\ \phi_{1} \end{pmatrix} }$,
	${\scriptstyle \begin{pmatrix} \psi_{1} \\ \psi_{2} \end{pmatrix} } \to {\scriptstyle \begin{pmatrix} - i \g_{1} \psi_{2} \\ - i \g_{1} \psi_{1} \end{pmatrix} }$,} \\ %2
	& $\sg \to + \sg$, &  \\ %3
	& $\la \to - i \g_{1} \la$, $\overline{\la} \to + i \g_{1} \overline{\la}$, & \multirow{2}{*}{${\scriptstyle \begin{pmatrix} F_{1} \\ F_{2} \end{pmatrix} } \to {\scriptstyle \begin{pmatrix} F_{2} \\ F_{1} \end{pmatrix} }$, \ with charges ${\scriptstyle \begin{pmatrix} + \qch \\ - \qch \end{pmatrix} }$.} \\ %4
	& $D \to - D$. & \\ \hline %5
	\end{tabular}
\end{align}

%%%%%%%%%%%%%%%%%% section 2.2 %%%%%%%%%%%%%%%%%%%%%%%%%%%%%%%%
\subsection{Localization}
%The partition function of the theory on $\RP \times \Sbb^1$ is called the SCI defined as the path integral over it \cite{Tanaka:2014oda}. We compute it exactly by the localization technique and here will list its formulas mainly necessary to describe Abelian mirror symmetry in the next section. See \cite{Tanaka:2014oda, Tanaka:2015pwa} for detailed calculations. To make the usage of the SCI systematic, we set the quiver rule for physical information about the theory.
The partition function of the theory on $\RP \times \Sbb^1$ is called the SCI defined as the path integral over it \cite{Tanaka:2014oda}. We compute it exactly by the localization technique and will list its formulas as the quiver rule for physical information, which are mainly necessary in the next section. See \cite{Tanaka:2014oda, Tanaka:2015pwa} for detailed calculations.
% For global symmetries
First of all, we draw $\xymatrix {*++[o][F.]{} }$ as a global symmetry with its background gauge field. The localized loci for the vector multiplet are given by

\begin{align} % Loci
\Pcal\tx{-type }
\xymatrix {*++[o][F.]{\text{\scriptsize$s^{\pm}, \theta$}} }
&:
A = s^{\pm} A_{\text{flat}} + \frac{\theta}{2 \pi} \de y, \quad
\sg =0, \quad
\lp \begin{array}{ll}
	s^{+} = 0  \\
	s^{-} = 1  
\end{array},
\quad
\theta \in [ 0, 2 \pi ]
\rp, \label{PBG} \\[.5em] %1
\Ccal\Pcal\tx{-type }
\xymatrix {*++[o][F.]{\text{\scriptsize$s, \theta_{\pm}$}} }
&:
A = s \ A_{\text{mon}} + \frac{\theta_\pm}{2 \pi} \de y, \quad
\sg = - s, \quad
\lp s \in \mathbb{Z}, \quad
	\begin{array}{ll}
	\theta_+ = 0  \\
	\theta_- = \pi
	\end{array}
\rp, \label{CPBG}
\end{align}
where $A_{\text{flat}}$ is the nontrivial flat connection on $\RP$, $A_{\text{mon}}$ is the Dirac monopole configuration with monopole charge $2$, and $\theta$ ($\theta_{\pm}$) is a Wilson line phase along $\Sbb^{1}$. The localization computation provides the following rules for the one-loop determinans of the matters with some R-charge $\Dl$ which are coupled to global symmetries with charges $\qch_{1}$ and $\qch_{2}$:
\begin{align}
&
\xymatrix{
*++[o][F.]{\text{\scriptsize$s_{1}^{\pm}, \theta_1$}}
&
*+[F-]{\Phi}
\ar@{.>}[l]^{\underbrace{}_{\qch_1}}
\ar@{.>}[r]^{}_{\underbrace{}_{\qch_2}}
&
*++[o][F.]{\text{\scriptsize$s_{2}^{\pm}, \theta_2$}}
}
=
\lc \hspace{-.35em} \begin{array}{ll}
	\lp q^{\frac{\Dl - 1}{8}} e^{\frac{i ( \qch_1 \theta_1 + \qch_2 \theta_2 )}{4}} \rp^{+1}
	\frac
	{( e^{- i ( \qch_1 \theta_1 + \qch_2 \theta_2 )} q^{\frac{2 - \Dl}{2}} ; q^{2} )_{\infty}}
	{( e^{ + i ( \qch_1 \theta_1 + \qch_2 \theta_2 )} q^{\frac{0 + \Dl}{2}} ; q^{2} )_{\infty}} 
	&
	\text{for } ( - 1 )^{\qch_1 s_{1}^{\pm} + \qch_2 s_{2}^{\pm}} = + 1, \\[.5em]
	\lp q^{\frac{\Dl - 1}{8}} e^{\frac{i ( \qch_1 \theta_1 + \qch_2 \theta_2 )}{4}} \rp^{-1}
	\frac
	{( e^{- i ( \qch_1 \theta_2 + \qch_2 \theta_2 )} q^{\frac{4 - \Dl}{2}} ; q^{2} )_{\infty}}
	{( e^{+ i ( \qch_1 \theta_1 + \qch_2 \theta_2 )} q^{\frac{2 + \Dl}{2}} ; q^{2} )_{\infty}}
	& 
	\text{for } ( - 1 )^{\qch_1 s_{1}^{\pm} + \qch_2 s_{2}^{\pm}} = - 1,
\end{array} \right. \label{s1loop} \\ %1, 2
&
\xymatrix{
*++[o][F.]{\text{\scriptsize$s_{1}^{\pm}, \theta_1$}} 
\ar@{<.}[r]<1mm>
\ar@{<.}[r]<-1mm> _{\underbrace{}_{\qch_1}}
&
*+[F-]{^{\Phi_1}_{\Phi_2}}
\ar@{.>}[r]<1mm>
\ar@{<.}[r]<-1mm> _{\underbrace{}_{\qch_2}}
&
*++[o][F.]{\text{\scriptsize$s_2, \theta_{2 \pm}$}}
}
=
\lp q^{\frac{1 - \Dl}{2}} e^{- i \qch_1 \theta_1} \rp^{| \qch_2 s_2 |}
\frac
{( e^{- i ( \qch_1 \theta{_1} +\qch_2 \theta{_2}{_\pm} )} q^{| \qch_2 s_2 | + \frac{2 - \Dl}{2}} ; q )_\infty}
{( e^{+ i ( \qch_1 \theta{_1} + \qch_2 \theta{_2}{_\pm} )} q^{| \qch_2 s_2 | + \frac{0 + \Dl}{2}} ; q )_\infty}.
\label{d1loop} %3
\end{align}
In the quivers, we depict the flavors by boxes, and a box with two matters $( \Phi_{1}, \Phi_{2} )$ implies $\Phi_{\tx{d}}$. The direction of the arrow corresponds to the sign of the charge, and, in the following, the number of arrowheads represents the absolute value of the charge for the symmetry.

% For gauge symmetries
Next, the recipes for gauge symmetries are given by gauging the global symmetries, that is, summing over the allowed sectors of their parameters. We use the following nodes $\xymatrix {*++[o][F-]{} }$ for a $\Pcal$-type and a $\Ccal\Pcal$-type gauge symmetry:
\begin{align}
\begin{xy}
{\ar@{->} (12.5,0) *++[o][F-]{\text{\scriptsize$\Pcal$}} }
\end{xy}
&=
q^{+ \frac{1}{8}} \frac{( q^2 ; q^2 )_\infty}{( q ; q^2 )_\infty}
\sum_{s^{\pm}}
\frac{1}{2 \pi}
\int_{0}^{2 \pi} \de \theta \
\lp
\begin{xy}
{\ar@{.>} (12.5,0) *++[o][F.]{\text{\scriptsize$s^{\pm}, \theta$}} }
\end{xy}
\rp, \label{Psum} \\ %1
\begin{xy}
\ar@{<-} (0,1);(8.5,1);
\ar@{->} (0,-1);(8.5,-1);
\ar@{} (11.5,0) *++[o][F-]{\text{\tiny$\Ccal\Pcal$}}
\end{xy}
&=
q^{- \frac{1}{8}} \frac{( q ; q^2 )_\infty}{( q^2 ; q^2 )_\infty}
\sum_{s \in \mathbb{Z}}
\frac{1}{2}
\sum_{\theta_\pm}
\lp
\begin{xy}
\ar@{<.} (0,1);(8.5,1);
\ar@{.>} (0,-1);(8.5,-1);
\ar@{} (13,0) *++[o][F.]{\text{\scriptsize$s , \theta_\pm$}}
\end{xy}
\rp. \label{CPsum}
\end{align}

% BF term
Finally, the Chern-Simons coupling $\frac{1}{2 \pi} \int B \w \hspace{.075em} \de A$ of two gauge fields $A$ and $B$ called a BF term is turned to be parity-even only when $A$ is $\Pcal$-type and $B$ is $\Ccal\Pcal$-type (or vice versa). Substituting the loci \eqref{PBG} and \eqref{CPBG} into the BF term leads to a quiver rule
\begin{align}
\xymatrix{
*++[o][F.]{\text{\scriptsize$s^{\pm}, \theta$}}
\ar @{.}[r]^{\text{BF}}
&
*++[o][F.]{\text{\scriptsize$s, \theta_\pm$}}
}
&=
e^{i s^{\pm} \theta{_\pm}}
e^{i s \theta}. \label{BFvalue}
\end{align}

%%%%%%%%%%%%%%%%%% section 3 %%%%%%%%%%%%%%%%%%%%%%%%%%%%%%%%%
\section{Abelian mirror symmetry}
The statement of 3d mirror symmetry is that distinct theories go down to the same IR fixed point along the RG flow. Here, we would like to show the $\Ncal = 2$ version of this duality with a single flavor, that is, the one between the SQED and the XYZ model. The SQED contains a vector multiplet and matter multiplets $Q$ and $\tilde{Q}$ charged under the U$(1)$ gauge symmetry in the opposite manner. The XYZ model consists only of three matter fields $X$, $Y$, and $Z$ coupled with each other through the superpotential $XYZ$.

One key ingredient for mirror symmetry is to match global symmetries; the SQED has a topological symmetry U$(1)_{J}$ and a flavor symmetry U$(1)_{A}$, which exactly corresponds to two flavor symmetries U$(1)_{V}$ and U$(1)_{A}$ in the XYZ model. This fact should be translated into the parameter identifications in the SCIs of two theories (see \cite{Tanaka:2015pwa} for their charge assignments).

Another important fact is the correspondence of the moduli spaces in these theories. The moduli in the SQED are casted by the scalar $\pm \sg$ for the Coulomb branch and the product $Q \tilde{Q}$ for the Higgs branch, and they are mapped into $( X, Y )$ and $Z$, respectively, as the moduli in the XYZ model. While we do not have a priori principle to fix the parity conditions of $X$, $Y$, and $Z$, the parity action on the matters in the SQED can be readily found with choosing the $\Pcal$-type or $\Ccal\Pcal$-type vector  multiplet. Thus, we apply the moduli correspondence to determining an appropriate parity condition in the XYZ model dual to the SQED having each of $V^{( \Pcal )}$ and $V^{( \Ccal\Pcal )}$ as done in \cite{Tanaka:2015pwa}. We should notice that there are two types of mirror symmetry on $\RP \times \Sbb^1$ depending on whether the dynamical vector multiplet in the SQED is ${\cal P}$-type or ${\cal CP}$-type. In the following, we call the former SQED$^{( \Pcal )}$ and the latter SQED$^{( \Ccal\Pcal )}$.

%%%%%%%%%%%%%%%%%% section 3.1 %%%%%%%%%%%%%%%%%%%%%%%%%%%%%%%%
\subsection{SQED$^{( \Pcal )}$ vs $^{\text{X}}_{\text{Y}}\text{Z}$}
%The left-hand sides of \eqref{quiverN2P+} and \eqref{mirrorN2P+} show the quiver diagram and the SCI of $\Pcal$-SQED. Its dual XYZ model is denoted by ${^\text{X}_\text{Y}}\text{Z}$ whose quiver on the right-hand side of \eqref{quiverN2P+} contains a doublet formed by $X$ and $Y$. % while $Z$ is an independent field.
%$\Ncal = 2$ mirror symmetry between these theories via the SCIs is realized as the equality \eqref{mirrorN2P+} under the identifications $a = \tilde{a}$ and $s = \tilde{s}$. The definition of $a$ ($\tilde{a}$) includes fugacities $e^{i \theta_{A}}$ ($e^{i \tilde{\theta}_{A}}$) for U$(1)_{A}$ in each theory, and $s$ ($\tilde{s}$) is a monopole flux for U$(1)_{J}$ (U$(1)_{V}$). We give the exact verification \cite{Tanaka:2014oda} achieved by the $q$-binomial theorem \cite{Gasper}.
The left-hand side of \eqref{quiverN2P+} and \eqref{mirrorN2P+} shows the quiver diagram and the index of SQED$^{( \Pcal )}$. Its dual XYZ model is denoted by ${^\text{X}_\text{Y}}\text{Z}$ whose quiver on the right-hand side of \eqref{quiverN2P+} contains a doublet formed by $X$ and $Y$ (represented by a single box). %while $Z$ is an independent field.
$\Ncal = 2$ mirror symmetry between these theories by means of the superconformal indices is realized as the equality \eqref{mirrorN2P+} under the identifications $a = \tilde{a}$, $s = \tilde{s}$, and $\theta_{J \pm} = \theta_{V \pm}$. Here, $a := e^{- 2 i \theta_{A}} q^{1 - \Dl}$ ($\tilde{a} :=  e^{2 i \tilde{\theta}_{A}} q^{1 - \Dl}$) includes a Wilson line phase for U$(1)_{A}$ in each theory. $s$ ($\tilde{s}$) and $\theta_{J \pm}$ ($\theta_{V \pm}$) are a monopole flux and a Wilson line phase for U$(1)_{J}$ (U$(1)_{V}$). We have given the exact verification and the mathematical generalization of \eqref{mirrorN2P+} in \cite{Tanaka:2014oda} by the $q$-binomial theorem \cite{Gasper}.

% N=2 qiuiver of P-SQED^+- and corresponding XYZ
%\vspace{-2em}
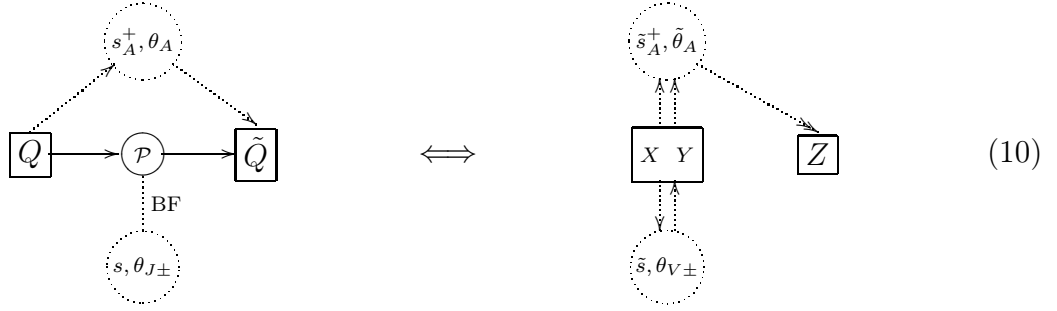
\begin{align} \label{quiverN2P+}
\begin{xy} % P-SQED^+-
\ar@{} (0,0) *+[F-]{\text{$Q$}},
\ar@{->} (2.5,0);(15,0) *++[o][F-]{\text{\scriptsize$\Pcal$}},
\ar@{->} (17.5,0);(30,0) *+[F-]{\text{$\tilde{Q}$}},
\ar@{<.} (0,3);(15,15) *++[o][F.]{\text{\scriptsize$s_{A}^{+}, \theta_A$}};
\ar@{.>} (19,12);(30,4),
\ar@{.}_{\text{BF}} (15,-3);(15,-10);
\ar@{} (15,-15) *++[o][F.]{\text{\scriptsize$s, \theta_{J \pm}$}}
\end{xy}
&&
\hspace{-2em}
\Longleftrightarrow
&&
\hspace{-2em}
\begin{xy} % XYZ
\ar@{} (0,0) *+[F-]{\text{\scriptsize$\begin{matrix} X & Y \end{matrix}$}};
\ar@{<.} (-1,3.5);(-1,10);
\ar@{<.} (1,3.5);(1,10);
\ar@{} (0,15) *++[o][F.]{\text{\scriptsize$\tilde{s}_{A}^{+}, \tilde{\theta}_A$}};
\ar@{<.} (-1,-3.5);(-1,-10);
\ar@{.>} (1,-3.5);(1,-10);
\ar@{} (0,-15) *++[o][F.]{\text{\scriptsize$\tilde{s}, \theta_{V \pm}$}},
\ar@{.>>} (20,3);(4,12);
\ar@{} (20,0) *+[F-]{\text{$Z$}},
\end{xy}
\end{align}

% Index for N=2 mirror sym
\begin{align} \label{mirrorN2P+}
&
q^{\frac{1}{8}}
\frac{( q^{2} ; q^{2} )_{\infty}}{( q ; q^{2} )_{\infty}}
\sum_{s^{\pm}}
\int_{0}^{2 \pi} \frac{d \theta}{2 \pi}
e^{i s \theta}
e^{i s^{\pm} \theta_{J \pm}}
\lc
a^{- \frac{1}{4} + \frac{s^{\pm}}{2}}
\frac
{( e^{- i \theta} a^{\frac{1}{2}} q^{\frac{1}{2} + s^{\pm}}, e^{i \theta} a^{\frac{1}{2}} q^{\frac{1}{2} + s^{\pm}}; q^{2} )_{\infty}}
{( e^{i \theta} a^{- \frac{1}{2}} q^{\frac{1}{2} + s^{\pm}}, e^{- i \theta} a^{- \frac{1}{2}} q^{\frac{1}{2} + s^{\pm}} ; q^{2} )_{\infty}}
\rc \notag \\ %1
&=
q^{\frac{1}{8}} \tilde{a}^{- \frac{1}{4}}
\left( \tilde{a}^{- \frac{1}{2}} q^{\frac{1}{2}} \right)^{| \tilde{s} |}
\frac
{( e^{i \theta_{V \pm}} \tilde{a}^{- \frac{1}{2}} q^{1 + |\tilde{s}|} ; q )_{\infty}}
{( e^{i \theta_{V \pm}}\tilde{a}^{\frac{1}{2}} q^{|\tilde{s}|} ; q )_{\infty}}
\frac{( \tilde{a} ; q^{2} )_{\infty}}{( \tilde{a}^{- 1} q ; q^{2} )_{\infty}}. %2
\\[-.5em] \notag
\end{align}

%%%%%%%%%%%%%%%%%% section 3.2 %%%%%%%%%%%%%%%%%%%%%%%%%%%%%%%%
\subsection{SQED$^{( \Ccal\Pcal )}$ vs XYZ}
For SQED$^{( {\cal CP} )}$ (the left quiver in \eqref{quiverN2CP+}), the charged matters $Q$ and $\tilde{Q}$ are combined into a doublet field (again by a single box with them), on the other hand, the dual theory witten by $\text{X}\text{Y}\text{Z}$ (the right quiver in \eqref{quiverN2CP+}) is comprised of three matter multiplets which behave as each single field on $\RP \times \Sbb^1$. $\mathcal{N} = 2$ mirror symmetry for these theories can be expressed by \eqref{mirrorN2CP+} under $a = \tilde{a}$,\footnote{We use the different definition $a := e^{2 i \theta_{A}} q^{\Dl}$ ($\tilde{a} := e^{- 2 i \tilde{\theta}_{A}} q^{\Dl}$) from the previous one.} $s^{\pm} = \tilde{s}^{\pm}$, and $w = \tilde{w}$, where $s^{\pm}$ ($\tilde{s}^{\pm}$) and  $w := e^{i \theta_{J}}$ ($\tilde{w} := e^{i \tilde{\theta}_{V}}$) are parameters associated with U$(1)_{J}$ (U$(1)_{V}$). As for the previous case, we have accomplished the complete proof of \eqref{mirrorN2CP+} in \cite{Tanaka:2015pwa} by applying the Ramanujan's summation formula \cite{Gasper}.
% N=2 qiuiver of CP-SQED^+- and corresponding XYZ
%\vspace{-2em}
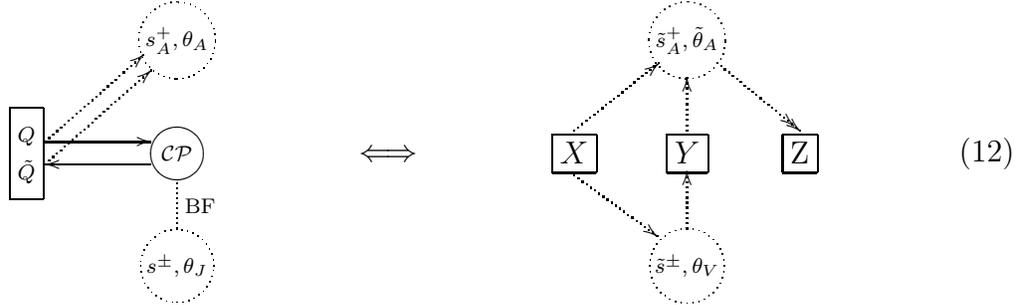
\begin{align} \label{quiverN2CP+}
\begin{xy} % CP-SQED^+-
\ar@{<.} (10,2);(22.5,13);
\ar@{<.} (10,-1);(23.5,11);
\ar@{} (7,0) *+[F-]{\text{\scriptsize$\begin{matrix} Q \\ \tilde{Q} \end{matrix}$}},
\ar@{<-} (9.5,1.5);(23.5,1.5);
\ar@{->} (9.5,-1.5);(23.5,-1.5);
\ar@{} (27,0) *++[o][F-]{\text{\scriptsize$\mathcal{CP}$}};
\ar@{} (27,15) *++[o][F.]{\text{\scriptsize$s_{A}^{+}, \theta_A$}};
\ar@{.}^{\text{BF}} (27,-10);(27,-4);
\ar@{} (27,-15) *++[o][F.]{\text{\scriptsize$s^{\pm}, \theta_{J}$}}
\end{xy}
&&
\hspace{-2em}
\Longleftrightarrow
&&
\hspace{-2em}
\begin{xy} % XYZ
\ar@{} (0,0) *+[F-]{\text{$X$}};
(15,0) *+[F-]{\text{$Y$}},
\ar@{} (15,0);(30,0) *+[F-]{\text{Z}},
\ar@{<.} (0,3);(15,15) *++[o][F.]{\text{\scriptsize$\tilde{s}_{A}^{+}, \tilde{\theta}_A$}};
\ar@{<.} (15,3);(15,10);
\ar@{.>>} (19,12);(30,3),
\ar@{.>} (15,-2.5);(15,-10);
\ar@{<.} (0,-3);(11,-11);
\ar@{} (15,-15) *++[o][F.]{\text{\scriptsize$\tilde{s}^{\pm}, \theta_{V}$}}
\end{xy}
\end{align}

% Index for N=2 mirror sym
\begin{align} \label{mirrorN2CP+}
&
q^{- \frac{1}{8}}
\frac{( q; q^{2} )_{\infty}}{( q^{2}; q^{2} )_{\infty}}
\sum_{s \in \Zbb}
\frac{1}{2}
\sum_{\theta_{\pm}}
e^{i s^{\pm} \theta_{\pm}}
w^{s}
\left( q^{\frac{1}{2}} a ^{- \frac{1}{2}} \right)^{| s |}
\frac{( e^{i \theta_{\pm}} a^{- \frac{1}{2}} q^{| s | + 1}; q )_{\infty}}{( e^{i \theta_{\pm}} a^{\frac{1}{2}} q^{| s |}; q )_{\infty}} \notag \\ %1
&=
q^{- \frac{1}{8}}
\tilde{a}^{\frac{s^{\pm}}{2}}
\frac
{( \tilde{a}^{\frac{1}{2}} \tilde{w}^{- 1} q^{\frac{1}{2} + \tilde{s}^{\pm}}, \tilde{a}^{\frac{1}{2}} \tilde{w} q^{\frac{1}{2}  + \tilde{s}^{\pm}}, \tilde{a}^{- 1} q; q^{2} )_{\infty}}
{( \tilde{a}^{- \frac{1}{2}} \tilde{w} q^{\frac{1}{2}  + \tilde{s}^{\pm}}, \tilde{a}^{- \frac{1}{2}} \tilde{w}^{- 1} q^{\frac{1}{2}  + \tilde{s}^{\pm}}, \tilde{a}; q^{2} )_{\infty}}. %2
\end{align}

%%%%%%%%%%%%%%%%%% section 4 %%%%%%%%%%%%%%%%%%%%%%%%%%%%%%%%%
\section{Summary}
We have established single-flavor Abelian mirror symmetry on $\RP \times \Sbb^1$ by exactly calculating the superconformal indices. Our mirror symmetry is actually turned to be quite non-trivial mathematical identities to which we could give the rigorous proof \cite{Tanaka:2014oda, Tanaka:2015pwa}. We can immediately construct the $\Ncal = 4$ version of mirror symmetry from $\Ncal = 2$ one, which provides fundamental formulas to generalize it to multi-flavor mirror symmetry. In addition, the operation $S$ of a SL$(2,\Zbb)$ group acting on 3d theories \cite{Witten:2003ya} can be consistently defined on $\RP \times \Sbb^1$ and is used to realize mirror symmetry with inserting loop operators. One can see the precise discussion of these extensions in \cite{Mori}.

%%%%%%%%%%%%%%%%%% Acknowledgments %%%%%%%%%%%%%%%%%%%%%%%%%%%%%
%\acknowledgments{We would like to thank Heng-Yu Chen, Koji Hashimoto, Kazuo Hosomichi, Norihiro Iizuka, Yosuke Imamura, Taro Kimura, Yousuke Ohyama, Yuji Sugawara, Masato Taki, Seiji Terashima, and Satoshi Yamaguchi. This research was supported by the RIKEN iTHES Project. The work of H.M. and T.M. was supported in part by the JSPS Research Fellowship for Young Scientists.}

%%%%%%%%%%%%%%%%%% References %%%%%%%%%%%%%%%%%%%%%%%%%%%%%%%%
%\bibliographystyle{maik}
%%\nocite{*}
%\bibliography{150716999QTSref}

\end{document}